\def\year{2021}\relax
\documentclass[letterpaper]{article} 
\usepackage{aaai21}  
\usepackage{times}  
\usepackage{helvet} 
\usepackage{courier}  
\usepackage[hyphens]{url}  
\usepackage{graphicx} 
\usepackage{natbib}  
\usepackage{caption} 
\title{Disambiguating Disinformation: Extending Beyond the Veracity of Online Content }
\author{
    Keeley Erhardt and
    Alex Pentland\
    \\
}
\affiliations{
    MIT Media Lab\\
    75 Amherst Street, Cambridge, MA 02139\\
    {\{keeley,pentland\}}@mit.edu
}

\begin{document}

\maketitle

\begin{abstract}
Following the 2016 US presidential election and the now overwhelming evidence of Russian interference, there has been an explosion of interest in the phenomenon of “fake news”. To date, research on false news has centered around detecting content from low-credibility sources and analyzing how this content spreads across online platforms. Misinformation poses clear risks, yet research agendas that overemphasize veracity miss the opportunity to truly understand the Kremlin-led disinformation campaign that shook so many Americans. In this paper, we present a definition for disinformation—\textit{a set or sequence of orchestrated, agenda-driven information actions with the intent to deceive}—that is useful in contextualizing Russian interference in 2016 and disinformation campaigns more broadly. We expand on our ongoing work to operationalize this definition and demonstrate how detecting disinformation must extend beyond assessing the credibility of a specific publisher, user, or story.
\end{abstract}

\section{Introduction}

\noindent The distinction between disinformation and, the more commonly identified, misinformation is in the underlying intention. Disinformation encompasses intentional efforts by state and non-state actors to manipulate public opinion and alter how individuals perceive events. This may be through agenda promotion—fostering support for a specific cause, message, or idea \cite{starbird2019disinformation}, or through efforts to confuse and distract, and thereby “kill the possibility of debate and a reality-based politics” \cite{pomerantsev2014menace}. In comparison, misinformation has no explicit objective and may be spread organically via people’s natural tendency to engage with novel claims \cite{vosoughi2018spread}. Disinformation is often defined as false information, deliberately designed to deceive. However, this definition fails to account for the etymology of the word, rooted in the Soviet practice of \textit{dezinformatsiya}—an explicitly orchestrated information action implemented in pursuit of a Soviet agenda. Furthermore, the definition does not clearly distinguish disinformation from “fake news”, similar, yet distinct, concepts.

Fundamentally, disinformation has three core elements: the \textit{intent to deceive}, \textit{orchestrated activity}, and an \textit{underlying agenda}. The intent to deceive may be expressed through deliberately false information \cite{kumar2016disinformation}, disguising one's true identity \cite{arif2018acting}, or the coordinated promotion of a hashtag, article, or storyline by inauthentic social accounts (e.g., bots) \cite{zannettou2019disinformation}. Beyond its deceptive nature, disinformation must also encompass a coordinated effort with an agenda, defined by Merriam-Webster as “an underlying, often ideological, plan or program”. Clickbait that contains unsubstantiated celebrity gossip is not disinformation because, though deceptive, there is no indication of orchestrated, agenda-driven behavior.

In comparison, Russian interference in 2016 was deceptive in nature; the Kremlin-linked Internet Research Agency (IRA) hired employees to pose as Americans, or even local news outlets, to share inflammatory images, and to organize “grassroots” events \cite{yin2018your,diresta2019tactics}. It was coordinated; the IRA was indicted for hiring 80 full-time employees to promote hashtags and articles on select topics, distributed by IRA managers at the beginning of shifts \cite{chen2015agency,doj2018indictment}. And, it was agenda-driven; according to the Intelligence Community Assessment report \textit{Assessing Russian Activities and Intentions in Recent US Elections}: “Russia’s goals were to undermine public faith in the US democratic process, denigrate Secretary Clinton, and harm her electability and potential presidency” \cite{activities2017intentions}.  

\section{Related Work and Challenges}

Research that emerged following the 2016 US presidential election has primarily focused on identifying what is now commonly referred to as “fake news”. Fake news and disinformation are closely related and therefore often conflated. Both are rooted in deception, but fake news is not necessarily promoted through orchestrated activity and might not have an underlying agenda. Previous work has approached and defined fake news in various ways:

\begin{itemize}
\item identifying three sub-types: “serious fabrications”, hoaxes, and satire \cite{rubin2015deception},
\item comparing the information to content produced by legitimate news outlets, similar in form but not in organizational process nor intent \cite{lazer2018science},
\item or expressing, more narrowly, that fake news is “a news article that is intentionally and verifiably false and could mislead readers” \cite{shu2017fake}.
\end{itemize}

These definitions highlight the deceptive nature of fake news, while omitting mention of coordinated or agenda-driven behavior. Despite the omission, we briefly review fake news detection approaches because there is substantial overlap in understanding the intent to deceive—e.g., false content, inauthentic users, and low-credibility publishers. We then describe the limitations inherent in current approaches to fake news identification, and the need for further research to better understand the coordinated and agenda-based elements of disinformation campaigns.

\subsection{Fake news detection and limitations}

Work to automate the detection of fake news can be broadly categorized as focusing on either the news media content, or the social media context \cite{shu2017fake}. Techniques to analyze the veracity of news articles commonly focus on the source, i.e. the publisher and/or the author(s) \cite{shao2018spread}, and the content. Reviewing the broader social media context allows one to also incorporate information on the user sharing the news content \cite{kumar2016disinformation}, and the response from other users in the network \cite{kwon2017rumor,rosenfeld2020kernel}. Yet, when analyzing how stories spread through social networks and how social media users engage with the content, researchers often rely on human labeling of the information or survey responses in small-scale studies \cite{castillo2011information,guess2019less}, or they simplify the problem by focusing on the credibility of the publisher \cite{lazer2018science,shao2018spread,bovet2019influence}. An article shared on social media is classified as credible or not credible based on whether the domain of the extracted URL (Uniform Resource Locator) is identified as “low-credibility” by a fact-checking or media bias organization.

This approach sheds light on the proliferation of unreliable, biased, or hyper partisan information, but should not be mistaken for disinformation detection. Studies have shown that of the 556 Twitter accounts found to be operated by IRA employees and weaponized in the months leading up to the 2016 US presidential election, the five most active accounts were fake news outlets, “two from Syria, including todayinsyria, an account purporting to be a local Berlin outlet, and two masquerading as U.S. local news outlets, namely DailySanFran and KansasDailynews” \cite{yin2018your}. These state-sponsored fake news outlets never made their way onto popular fake news watch lists \cite{shao2018spread}, potentially because they never attracted much attention \cite{yin2018your}. Even well-maintained fact-checking and media bias resources cannot stay entirely up-to-date and will never be able to assess all domains that publish news or news-like content. Popular reference lists that categorize domains as biased, conspiratorial, satirical, etc. typically include less than a thousand sources. These lists under count large numbers of more niche, emerging, or transient low-credibility domains.

\subsection{Coordinated behavior}

As identified by Starbird et al., “distinguishing between orchestrated, explicitly coordinated, information operations and the emergent, organic behaviors of an online crowd” is a critical challenge that remains understudied \cite{starbird2019disinformation}. Disinformation campaigns consist of numerous information actions over a period of time, so moving beyond identifying if a single social media post or news article is false, misleading, or inaccurate requires drawing threads between these individual pieces of content, and the sources sharing them. However, understanding which online accounts are controlled by the same offline entity is challenging without access to user data not readily available to researchers (often due to privacy concerns), such as the email address or phone number associated with an account. Similarly, metadata that could be useful in determining if seemingly unrelated posts might be part of the same campaign, such as the IP address that a post comes from, is rarely publicly available—resulting in an overemphasis on analyzing the data that is available (e.g., content) in academic work. 

Social media platforms themselves have access to more complete data. Beginning in October 2018, Twitter has published an archive of suspected state-linked information operations identified on the Twitter platform. Similarly, since 2018, Facebook had published a monthly Coordinated Inauthentic Behavior Report highlighting the Facebook Pages, Facebook Groups, and Facebook and Instagram accounts that they have taken down for engaging in “coordinated inauthentic behavior”. The behavior is defined in the reports as “coordinated efforts to manipulate public debate for a strategic goal where fake accounts are central to the operation”. This definition encompasses elements of deception (fake accounts), orchestration (coordinated efforts), and agenda-driven behavior (a strategic goal). These reports demonstrate that detection of inauthentic coordinated activity is possible with sufficient information—in February 2021 alone, networks from Thailand, Iran, Morocco, and Russia were dismantled, totaling 915 Facebook accounts, 606 Instagram accounts, 86 Facebook Pages, and 21 Facebook Groups \cite{facebook2021cib}. In the same month, Twitter published a dataset of 373 accounts attributed to state-linked campaigns originating from Iran, Armenia, and Russia \cite{twitter2021transparency}.

Despite this success, there is a clear gap in automated disinformation detection across platforms owned by different entities where there are more barriers to data sharing. Unsurprisingly, platforms typically aim to detect problematic users and content on their own platforms. Unfortunately, this bias is also represented in academic work that most commonly studies a single social network, despite the cross-platform nature of sophisticated influence campaigns \cite{diresta2019tactics}. This results in a single platform focus that inadequately addresses the coordinated dimension of disinformation. Shifting focus to cross-platform analysis will take concerted effort given the data sharing requirements, the increased time and effort required to normalize data from multiple social platforms, and the complexity introduced when developing models that perform well on data from different sources. 

\subsection{Agenda-driven information operations}

Discerning an underlying agenda is arguably even harder than detecting coordinated activity, or false information. Accurately assessing whether content furthers a particular message or idea requires an understanding of all possible agendas that an actor may have. To simplify the problem, one can instead attempt to detect whether a news article or social media post uses propaganda techniques. Propaganda is commonly defined as information that is not impartial and used to promote a particular view. Though detecting usage of propaganda techniques will not expose the specific agenda being promoted, it does provide insight into the likelihood that a piece of content is agenda-driven. Recent work has focused on detecting the use of propaganda in online media, in particular, the SemEval-2020 Task 11: Detection of Propaganda Techniques in News Articles. The task involved spotting “specific text fragments containing propaganda” and determining the propaganda technique used \cite{da2020semeval}. 

In analyzing agenda-driven behavior, it is important to highlight related activity that we do not classify as disinformation. For example, the Voice of America Extremism Watch Desk, launched in 2015, is explicitly designed to support an anti-extremism agenda. The desk produces content on extremism and terrorism, initially focused on the Islamic State and Al-Qaeda, and now expanded to other extremist groups. The coverage is coordinated and agenda-driven, but relies on persuasion more than propaganda and is intended to influence, not outright deceive. In comparison, the Chinese government’s efforts to discredit the Hong Kong protests in support of democracy \cite{myers2019china}, and Russian information operations during the Syrian civil war \cite{alami2018russia}, involved inauthentic social accounts and the promotion of misleading content. This intent to deceive and the tactics used to persuade differentiate these, otherwise very similar, influence efforts.

\section{Automated Disinformation Detection}

We are developing a system for detecting disinformation that includes five stages:

\begin{enumerate}
\item Data source selection, collection, and normalization
\item Detection and tracking of narratives
\item Classification of narratives as organic or orchestrated and inauthentic
\item Attribution
\item Measurement of the impact of a disinformation campaign
\end{enumerate}

\subsection{Data source selection, collection, and normalization}

We focus data collection on news media and social media content. News media can be collected directly from the publishing platform, or through following URLs shared on social media to the source content. We have started with extracting URLs from posts shared on social media and following these links to off-platform stories. This approach has a couple of advantages: we know social engagement data exists (because the story was discovered on a social platform), and we are not biasing the sample of stories by pre-selecting news outlets to pull stories from (reducing the likelihood that we miss content from niche blogs, fora, and transient websites). That said, there remains a bias due to the social platforms selected.

To mitigate this bias, we ingest data from a number of disparate social media platforms, including mainstream platforms and niche sources. This helps to ensure that the extracted URLs are a reasonable sample of the online information environment that individuals engage with. Monitoring a range of social platforms further enables us to better account for the cross-platform nature of sophisticated information operations, though it requires thoughtful data normalization to enable analysis across data types.

\subsection{Detection and tracking of narratives}

We identify narratives as the core unit of analysis, rather than users, publishers, or stories. We define a narrative similarly to Chambers et al., as a coherent set or sequence of events \cite{chambers2009unsupervised}, and define an event by following Dou et al.: “An occurrence causing change in the volume of text data that discusses the associated topic at a specific time… often associated with entities such as people and locations” \cite{dou2012event}. Rather than study how users engage with an individual story or publisher, we propose instead focusing on identifying the narratives that social media users are engaging with.

A narrative can be constructed through detecting, disambiguating, and clustering entities, topics, and events. In 2019, Fedoryszak et al. published  a paper on the Twitter real-time event detection system that, at a high-level, extracted entities from tweets, computed the co-occurrences of the entities to measure similarity, constructed a graph with entities as nodes and co-occurrence similarity as edge weight, and then clustered entities. By linking these clusters across time steps Fedoryszak et al. demonstrated the ability to detect and track the evolution of events using data from the complete Twitter Firehose \cite{fedoryszak2019real}. Other novel techniques have instead incrementally clustered tweets to understand narrative evolution \cite{hasan2019real}. We hypothesize that an approach focused on narratives supports more sophisticated disinformation detection, and enables a researcher to better account for the coordinated nature of disinformation campaigns, spanning many users, publishers, and stories.

\subsection{Classification of narratives as organic or orchestrated and inauthentic}

After assembling narratives, our ongoing work aims to classify them as organic (credible content with authentic social engagement), or orchestrated and inauthentic (low-credibility content and/or engagement from coordinated sources). As previously highlighted, this classification requires analyzing narratives across three criteria (\textit{intent to deceive}, \textit{orchestrated activity}, and an \textit{underlying agenda}). The features for this classification task can be based on a number of attributes, including the assessed credibility of the content, users, and/or publishers engaged in a narrative, indications of coordinated behavior (e.g., co-occurrence of highly similar posts), and signs of agenda-driven messaging (e.g., the use of propaganda techniques in news articles or social media posts).

These features span the news content and the social media context, and both are relevant to detecting whether there is an orchestrated effort to alter the information environment. Previous work into the social context of information disorders covers a wide range of user/account-, metadata-, content-, temporal-, and structural-based features \cite{castillo2011information,shu2017fake,kwon2017rumor,rosenfeld2020kernel}. Examples include account creation time (user), number of friends, following, and followers (user), post creation time (metadata), geolocation of post (metadata), occurrence of arbitrary words (content), time series of narrative related entities and events (temporal), time series of user engagement with a narrative (temporal), accounts friended, followed, and following (structural), and the network of accounts sharing narrative related content (structural). Similar content-based features can be used for detecting misleading and deceptive news stories, as well as agenda-driven news outlets.

\subsection{Attribution}

Building an understanding of online information operations has value in of itself, but without attempting to attribute these campaigns back to the actor responsible, key elements of the operation remain hidden. Attribution provides information that is useful in preventing similar campaigns from taking place in the future (through understanding the techniques preferred by a given actor), and it helps elucidate underlying motivations—useful in contextualizing information operations within the broader geopolitical climate. The information required to confidently make attribution assessments might be unavailable to most researchers, but we are exploring features that might provide clues about the sponsors of adversarial information actions.

\subsection{Measurement of the impact of a disinformation campaign}

And finally, after understanding the actor behind a campaign and their motivations, it becomes possible to begin to measure the effectiveness of a coordinated disinformation effort. Measuring the impact of an information operation has substantial overlap with the work done by Advertising Technology (AdTech) companies to understand the success of a marketing campaign. Social media companies, in particular, provide robust infrastructure for marketers to understand how their content performs across different audiences. Many of these platforms provide metrics on ad performance (e.g., reach), engagement (e.g., likes), and conversions (e.g., website views). Leveraging this same advertising infrastructure, political operatives can track how their influence campaigns are performing, and the platforms themselves can retroactively measure how many users were shown content later attributed to a disinformation operation. Still, online exposure does not necessarily correspond to offline behavioral changes, so the limitations of these measures require further study.

\section{Conclusion}

A state actor meddling in another country’s democratic processes is not novel \cite{prados2006safe}, though the interference is particularly salient when it takes on a new online element—leveraging popular social media platforms \cite{diresta2019tactics}, driving the creation of bogus online news sites that resemble the mainstream media \cite{mejias2017disinformation}, and manifesting inauthentic social accounts that take on imagined identities. To tackle these challenges, researchers must push beyond a fake news focus and strive to understand the broader online information ecosystem (outside of a limited subset of social media platforms and news outlets), the credibility of narratives, rather than the credibility of individual stories, publishers, social media posts, and social accounts, and the strategies pursued by state and non-state actors that choose to engage in adversarial information operations. We are actively in the process of developing the system outlined above, and are hopeful that this approach will enable more robust, sophisticated, and complete disinformation detection.

\bibliography{refs}

\end{document}